\documentstyle[12pt]{article}
\begin{document}

\begin{titlepage}

\title{Self-energy of Heavy Quark}

\vspace{0.5cm}

\author{ M. A. Ivanov\thanks{Permanent address: Bogoliubov Laboratory of
Theoretical Physics, Joint Institute for Nuclear Research,
141980 Dubna (Moscow region), Russia}
\,  and T. Mizutani\\
\\
\\
Department of Physics \\
Virginia Polytechnic Institute and State University\\
Blacksburg, VA 24061}
\maketitle

\vspace{1cm}

\abstract

We demonstrate that to calculate the self-energy of a heavy quark in the heavy
quark limit (or the heavy fermion limit in what is called the Baryon Chiral
Perturbation Theory), the use of standard dimensional regularization provides
only the quantum limit: opposite to the heavy quark (or classical) limit that
one wishes to obtain. We thus devised a standard ultraviolet/infrared
regularization procedure in calculating the one- and two-loop contributions to
the heavy quark self-energy in this limit. Then the one-loop result is shown to
provide the standard classical Coulomb self-energy of a static colour source
that is  linearly proportional to the untraviolet cutoff $\Lambda$. All the
two-loop contributions are found to be proportional to
$\Lambda \ln(\Lambda/\lambda)$
where $\lambda$ is the infrared cutoff. Often only the contribution from the
bubble (light quarks, gluon and ghost) insertion to the gluon propagator
has been considered as the $O(\alpha_s)$ correction to the Coulomb energy
to this order. Our result shows that other contributions are of the
same magnitude, thus have to be taken into account.

\end{titlepage}

\section{Introduction}
\baselineskip 20pt

Recently, many  discussions have been devoted to the definition of the of heavy
quark mass. This quantity is the basic parameter in the heavy quark physics.
Generally (not only for the heavy quarks in the heavy quark limit in which we
are interested, in the following) one of the most popular choices within the
perturbative scheme is the pole mass. In reference \cite{Tar} it has been shown
that the pole mass is gauge independent and infrared finite at the two-loop
level. It was checked at the three-loop level in \cite{GBGS} where the relation
between the mass of the modified minimal subtraction scheme
and the scheme-independent pole mass was also obtained. Later authors of work
\cite{BSUV} argue that precise definition of the pole mass in the heavy quark
limit may not be given once nonperturbative effects are included (a more
analytical
approach leading to the identical conclusion has been given independently in
\cite{BB}).
To demonstrate that, they adopted the standard effective running constant
(as a typical non-perturbative quantity) obtained from the summation of leading
logarithms to calculate the Coulomb energy of a static (or infinitely heavy)
colour source. This has produced an auxiliary singularity for small momentum
(the so-called, {\it infrared renormalon}) which, they claimed, leads to
the uncertainty of order $\Lambda_{QCD}/m_Q$ in the definition of the pole
mass.
An important observation which served as one of the bases to this demonstrative
calculation is that the widely used dimensional regularization may not be used
to analize self-energy diagrams in the heavy quark limit, in particular, for
separating the infrared and ultraviolet domains. The best example, as we shall
also present in the following section, is that this regularization method
leaves out the linear divergence that defines the self-energy of a static
quark in the lowest order: {\it the Coulomb energy}. Also it should be
emphasized here that the
{\sl Heavy Quark Limit} does make sense only after introducing a scale
parameter to the theory with which the quark mass may be compared, and there
is something awkward about it in the dimensional regularization method.

It may be useful to point out here that a quite analogous situation to
the heavy quark limit can be found in the
problem of the classical limit $\hbar\to 0$ for the self-energy of an electron
in QED \cite{EIM}. It was found that in the limit  $\hbar\to 0$, or more
precisely for the electron Compton wave length $\hbar/mc\ll r_0$, where
parameter $r_0$ regularizes small distances, the energy of the Coulomb field
by a change of radius $r_0$ is reproduced in the second order of perturabative
theory whereas the higher order contributions  were shown to vanish in this
limit.

Here we first study the quantum and classical limits for the lowest
order contributions to the quark self-energy and quark-gluon vertex, and
show that the standard dimensional regularization method only reproduces the
quantum limit. Since one may clearly identify the classical limit to be
equal to the currently fashonable heavy quark limit, we could conclude that
the dimensional regularization method is not suited for studying this case.
Then in the following section we calculate the two-loop contributions to
the quark self-energy in the heavy quark limit, exploiting the conventional
ultraviolet and infrared regularization method. We have found that each
contribution to this order is all proportional to
$\Lambda \ln{(\Lambda/\lambda)}$ where
$\Lambda$ and $\lambda$ are the ultraviolet and infrared cutoffs, respectively
and that, unlike in QED, these contributions do not mutually cancel out. We
point out that the gluonic bubble diagrams (Fig.4d) (implicitly) taken often
as the majour contribution in some processes is of the same magnitude as those
from vertex and quark self-energy diagrams (Figs. 4a-4c) for the self-energy
in the heavy quark limit. In the last section some comments will be given
regarding the alleged infrared renormalon and the Baryon (or heavy fermion)
Chiral Perturbation Theory.

\section{One-loop diagram}
\baselineskip 20pt

The self-energy of quark with mass {\bf m} in the second order of perturbative
theory is defined by the diagram in Fig.1,

\begin{equation}
\delta m_2=\Sigma_2(m)=C_Fg_s^2\int\!\!{d^4k\over (2\pi)^4i}
{1\over (-k^2)} \gamma^\mu {1\over m-\not\! k-\not\! p}\gamma_\mu |_{\not p=m}
\end{equation}
where $C_F=(N_c^2-1)/2N_c$, for a gauge group $SU(N_c)$.
First, we note that this expression coincides with the
electromagnetic self-energy except for the trivial factor $C_Fg_s^2$.
While the integral has no infrared divergence, it develops the linear
ultraviolet divergence. We are interested in this quantity in the heavy quark
limit: $m\to\infty$. As was mentioned in \cite{BSUV}, the dimensional
regularization cannot serve this purpose. To make this point  more clear,
we calculate the quark self-energy by way of introducing an ultraviolet
regularization,

\begin{equation}
{1\over (-k^2)}\to{F(-k^2/\Lambda^2)\over (-k^2)}.
\end{equation}
Here, the function $F(-k^2/\Lambda^2)$ is assumed to decrease rapidly in
the Euclidean region $k^2_E=-k^2\to +\infty$ providing the ultraviolet
convergence of the Feynman integrals. Now, the heavy quark limit (or classical
limit) means the case where $m\gg \Lambda$ in comparison with the quantum
limit: $m\ll \Lambda$.
The both limits may be realized using the Mellin representation for the
form factor $F(-k^2/\Lambda^2)$ under calculation of the integral in eq. (1).
We can write in the Euclidean region

\begin{equation}
F(k^2_E)=\sum\limits_{n=0}^\infty c_n(-)^n(k^2_E)^n=
{1\over 2i}\int\limits_{-\delta+i\infty}^
{-\delta-i\infty}{d\zeta\over\sin{\pi\zeta}} c(\zeta) (k^2_E)^\zeta,
\end{equation}
where $c(n)\equiv c_n$ for {\bf n} being positive integers, and $0<\delta<0.5$.

Finally, we have

\begin{eqnarray}
\delta m_2&=&
C_Fg_s^2
{1\over 2i} \int\limits_{-\delta+i\infty}^{-\delta-i\infty}
{d\zeta\over\sin{\pi\zeta}} c(\zeta) {1\over \Lambda^{2\zeta}}
\int\!\!{d^4k\over (2\pi)^4i}
{4m-2(\not\! k+\not\! p) \over (-k^2)^{1-\zeta}(m^2-(k+p)^2)} \nonumber\\
& & \nonumber\\
&=&C_Fg_s^2m
{1\over 2i}\int\limits_{-\delta+i\infty}^{-\delta-i\infty}
{d\zeta\over\sin{\pi\zeta}} c(\zeta) {1\over \Lambda^{2\zeta}}
{\Gamma(2-\zeta)\over\Gamma(1-\zeta)} \int\limits_0^1\!\! d\alpha
(1-\alpha)^{-\zeta} \int\!\!{d^4k\over (2\pi)^4i}
{4-2(1-\alpha)\over (\alpha^2 m^2-k^2)^{2-\zeta} } \nonumber\\
& &\nonumber\\
&=&{3\alpha_s\over 2\pi} C_F m
{1\over 2i}\int\limits_{-\delta+i\infty}^{-\delta-i\infty}
{d\zeta\over\sin{\pi\zeta}} c(\zeta)
{\Gamma(1+2\zeta)\over (-\zeta)}
\biggl({m^2\over \Lambda^2}\biggr)^\zeta
{\Gamma(1-\zeta)(1+\zeta)\over\Gamma(3+\zeta)}.
\end{eqnarray}

To get the leading term in $m/\Lambda$ in the quantum limit, we move
the integration counter to the right and take into account of the first
double pole at $\zeta=0$. Assuming that $F(0)=1$, we have

\begin{equation}
\delta m_2^{\rm qu}={3\alpha_s\over 4\pi} C_F m
\biggl\{\ln{\Lambda^2\over m^2}+O(1)\biggr\}.
\end{equation}

To get the leading term in $\Lambda/m$ in the classical limit, we move
the integration counter to the left and take into account the first
pole at $\zeta=-0.5$. From the standard relation between the Mellin and its
inverse transformations

$$c(-0.5)=2\int\limits_0^\infty\!\! dtF(t^2),$$
we have

\begin{equation}
\delta m_2^{\rm cl}={\alpha_s\over 2\pi} C_F \Lambda c(-0.5)
={\alpha_s\over \pi}C_F\Lambda \int\limits_0^\infty\!\! dt F(t^2).
\end{equation}

If we compare the above result with the calculation of the expression in Eq.(1)
devising the dimensional regularization,

\begin{equation}
\delta m_2={3\alpha_s\over 4\pi} C_F m
\biggl\{N_\epsilon+{5\over 3}-\ln{m^2\over\mu^2}\biggr\}
\end{equation}
where $N_\epsilon=1/\epsilon+\ln{4\pi}-\gamma$, ($\gamma$: the Euler-Mascheroni
constant) and $\mu$ is the subtraction
point, one can see that it only reproduces the quantum limit Eq.(5) for the
quark self-energy by identifying $\mu$ as the ultraviolet cutoff. Thus, the
naive dimensional regularization does not allow
to see the heavy quark limit.

Ordinarily, studies of heavy quark limit may be carried out by using the
expansion of quark propagator in the inverse quark mass:

\begin{eqnarray}
S(\not\! p+\not\! k)|_{\;\not p=m}&=&
i{M_0+\not\! k/2m \over
  [1+i/2m +f(\vec k^2/m^2)][k_4+2mif(\vec k^2/m^2)]}\nonumber\\
&& \nonumber\\
&=&{iM_0\over k_4+i\epsilon}+O({1\over m})
\end{eqnarray}
Here, $k_4=-ik_0$, $M_0=(1/2)(I+\gamma^0)$, and $f(x)=(1/2)(\sqrt{1+x}-1)$.
It is readily seen that the Dirac spinor $u(\vec p)$ for quark may be chosen as

\[
u(\vec p)=\left(\begin{array}{c}
                   I \\ 0
                  \end{array}
\right)
\]
so that $\bar u(\vec p)u(\vec p)$=$\bar u(\vec p)\gamma^0u(\vec p)$=1,
and $\bar u(\vec p)\vec\gamma u(\vec p)$=0.

In the second order: Eq.(1), upon introducing the ultraviolet regulator
to the otherwise linearly divergent expression, the leading contribution reads

\begin{equation}
\delta m_2^{\rm cl}=\Sigma_2(m)=
2\pi\alpha_s C_F \Lambda \int\!\!{d^3\vec k\over (2\pi)^3}
{F(\vec k^2)\over \vec k^2}
\end{equation}
which obviously coincides with the Eq.(6). One can see that this expression
is the classical Coulomb energy of a static color source.
To give more physical meaning to Eq.(9), one may introduce the charge density
defined  as

\begin{equation}
\rho(\vec r^{\; 2})=\int\!\!{d^3\vec k\over (2\pi)^3} e^{i\vec k\vec r}
\sqrt{F(\vec k^2)}
\end{equation}
supposing that $F(\vec k^2)>0$, and the electron (quark) "radius"
$ r_0=1/\Lambda$.

Then, we have

\begin{equation}
\delta m_2^{\rm cl}=
{g_s^2\over 2r_0} C_F
\int\!\!\int\!\! d\vec r_1 d\vec r_2 \rho(\vec r_1^{\, 2})
\left[{1\over 4\pi} {1\over |\vec r_1-\vec r_2|} \right] \rho(\vec r_2^{\, 2})
\end{equation}
in accordance with the classical result.

As was mentioned above, there is no infrared divergence in the second
order, but it plays the crucial role in higher orders in perturbative
expansion.

To understand its role, we compute the quark-gluon vertex induced by
both quark-gluon and self-gluon interactions: Figs.2-3. For simplicity we omit
all the common coefficients and color factors.

\medskip
\noindent
{\bf (a) Quark-gluon interaction (diagram in Fig.2):}

By introducing the gluon mass $\lambda$ serving to regularize the infrared
singularity we get

\begin{equation}
\Lambda_{1}^\nu(p,p)=\int\!\!{d^4k\over (2\pi)^4 i} {1\over \lambda^2-k^2}
{\gamma^\mu(m+\not\! k+\not\! p)\gamma^\nu
           (m+\not\! k+\not\! p)\gamma_\mu \over
\biggl(m^2-(k+p)^2 \biggr)^2 }|_{\not p=m}
\end{equation}

Using the form factor as in Eq. (2) one obtains

\[\Lambda_1^\nu(p,p)=\left\{ \begin{array}{ll}
\gamma^\nu{1\over(4\pi)^2} \biggl\{2\ln{\Lambda^2\over\lambda^2}
-\ln{\Lambda^2\over m^2}+O(1) \biggr\}  & \mbox{if $\Lambda\gg m$ }\\
 & \\
\gamma^\nu{1\over(4\pi)^2} \biggl\{2\ln{\Lambda^2\over\lambda^2}+O(1)
\biggr\}  & \mbox{if $\Lambda\ll m$ }
\end{array}
\right. \]
One can see that the infrared singularity defines the behavior of
the vertex in the heavy quark limit. Again result obtained from
the dimensional regularization for ultraviolet divergence corresponds only to
the quantum case:

\begin{equation}
\Lambda_1^\nu(p,p)=\gamma^\nu{1\over(4\pi)^2}
\biggl\{2\ln{\Lambda^2\over\lambda^2}
-N_\epsilon+\ln{m^2\over\mu^2}+O(1) \biggr\}.
\end{equation}
Note that if  we do not introduce the infrared cutoff within the dimensional
regularization scheme, we find
the auxiliary $1/\epsilon$ pole in the expression for the vertex in Eq.(12)
leading to

$$
\Lambda_1^\nu(p,p)=\gamma^\nu{1\over(4\pi)^2}
\biggl\{-2N_\epsilon^{IR}
-N_\epsilon^{UV}+3\ln{m^2\over\mu^2}+O(1) \biggr\}.
$$

\medskip
\noindent
{\bf (b) Self-gluon interaction (diagram in Fig.3):}

This contribution may be written as

\begin{equation}
\Lambda_{2}^\nu(p,p)=\int\!\!{d^4k\over (2\pi)^4 i} {1\over (\lambda^2-k^2)^2}
{-2k^2\gamma^\nu-4k^\nu \not\! k+2m k^\nu-\not\! k \not\! p \gamma^\nu
-\gamma^\nu \not\! p \not\! k  \over
\biggl(m^2-(k+p)^2 \biggr)^2 }|_{\not p=m}
\end{equation}

This integral turns out to be free from infrared singularity. Using the
ultraviolet form factor gives

\[\Lambda_2^\nu(p,p)=\left\{ \begin{array}{ll}
\gamma^\nu{1\over(4\pi)^2} \biggl\{
3\ln{\Lambda^2\over m^2}+O(1) \biggr\}  & \mbox{if $\Lambda\gg m$ }\\
 & \\
O({\Lambda\over m})  & \mbox{if $\Lambda\ll m$ }
\end{array}
\right. \]
It is readily seen that this diagram gives no contribution in the heavy quark
limit. Again result obtained from
the dimensional regularization for ultraviolet divergence corresponds only to
the quantum case:

\begin{equation}
\Lambda_2^\nu(p,p)=\gamma^\nu{1\over(4\pi)^2}
\biggl\{
3N_\epsilon-3\ln{m^2\over\mu^2}+O(1) \biggr\}.
\end{equation}

\section{Two-loop Diagrams in the Heavy Quark Limit}
\baselineskip 20pt

The contributions to the quark self-energy from the two-loops (or fourth
order in coupling constant) are defined by the diagrams in Fig.4.
As a consequence of our discussion in the previous section, we shall not adopt
the dimensional regularization, but merely use the ultraviolet form factor
Eq.(2)
together with the infrared regularization of the gluon propagator in the
following.

\newpage
\noindent
{\bf (a) Diagram in Fig. 4a.}

With $\Lambda$ and $\lambda$ being the ultraviolet and infrared cutoff masses,
as in the previous section, one finds

\begin{equation}
\delta m_4^{(a)}=\Sigma_4^{(a)}(m)=
C_4^{(a)}g^4_s\Lambda I_4^{(a)}(\kappa^2)
\end{equation}
where $\kappa^2\equiv\lambda^2/\Lambda^2$; $C_4^{(a)}=1$ (for QED), and
$t^a t^b t^a t^b=-C_F/2N_c$  (for QCD). The integral
$I_4^{(a)}(\kappa^2)$ is equal to

\begin{equation}
I_4^{(a)}(\kappa^2)=i\int\limits_{(E)}\!\!{d^4k_1\over (2\pi)^4}
\!\!\int\limits_{(E)}\!\!{d^4k_2\over (2\pi)^4}
{F(k_1^2)\over k_1^2+\kappa^2}{F(k_2^2)\over k_2^2+\kappa^2}
{1\over (k_{14}+i\epsilon)}{1\over (k_{14}+k_{24}+i\epsilon)}
{1\over (k_{24}+i\epsilon)}
\end{equation}
Here, to be definite,  the integration is over the dimensionless Euclidean
variables denoted as $(E)$ under the integration sign. The calculation
of this integral is given in Appendix. We have

\begin{equation}
\delta m_4^{(a)}=
2\pi\alpha_s C_F \Lambda \int\!\!{d^3\vec k\over (2\pi)^3}
{F(\vec k^2)\over \vec k^2}
\biggl[-{\alpha_s\over 4\pi} {1\over N_c} \ln(1/\kappa^2) \biggr].
\end{equation}

\noindent
{\bf (b) Diagram in Fig. 4b.}

Here one obtains

\begin{equation}
\delta m_4^{(b)}=\Sigma_4^{(b)}(m)=
C_4^{(b)}g^4_s\Lambda I_4^{(b)}(\kappa^2)
\end{equation}
where $C_4^{(b)}=-if^{abc} t^a t^b t^c=(N_c/2)C_F$ (of course
there is no QED counterpart to this diagram).
The integral $I_4^{(b)}(\kappa^2)$ is given as

\begin{eqnarray}
I_4^{(b)}(\kappa^2)&=&
\int\limits_{(E)}\!\!{d^4k_1\over (2\pi)^4}
\!\!\int\limits_{(E)}\!\!{d^4k_2\over (2\pi)^4}
\!\!\int\limits_{(E)}\!\!{d^4k_3\over (2\pi)^4}
{F(k_1^2)\over k_1^2+\kappa^2}{F(k_2^2)\over k_2^2+\kappa^2}
{F((k_1-k_2)^2)\over (k_1-k_2)^2+\kappa^2}\nonumber\\
& &\nonumber\\
&&\cdot{N_b\over (k_{14}+i\epsilon)(k_{24}+i\epsilon)}
\end{eqnarray}
with the numerator $N_b$ equal to

$$N_b=(\not\! k_1-2\not\! k_2)M_0\gamma^\nu M_0\gamma^\nu
+\gamma^\nu M_0(\not\! k_1+\not\! k_2)M_0\gamma^\nu+
\gamma^\nu M_0\gamma^\nu M_0 (\not\! k_2-2\not\! k_1).$$
Since the integration is over Euclidean variables here,
$\not\! k=\gamma^0 ik_4-\vec\gamma\vec k$. With this it is easy to check that
$\bar u(\vec p)N_bu(\vec p)=0$, which means that the contribution from
the diagram in the Fig.4b behaves as $\Lambda/m$ for large $m$, thus
vanishes in the heavy quark limit.

\medskip
\noindent
{\bf (c) Diagram in Fig. 4c.}

This contribution may be written as

\begin{equation}
\delta m_4^{(c)}=\Sigma_4^{(c)}(m)=
C_4^{(c)}g^4_s\Lambda I_4^{(c)}(\kappa^2)
\end{equation}
where $C_4^{(c)}=1$ (for QED), and $t^a t^a t^b t^b=C_F^2$ (for QCD).
The integral $I_4^{(c)}(\kappa^2)$ is equal to

\begin{eqnarray}
I_4^{(c)}(\kappa^2)&=&i\int\limits_{(E)}\!\!{d^4k_1\over (2\pi)^4}
\!\!\int\limits_{(E)}\!\!{d^4k_2\over (2\pi)^4}
{F(k_1^2)\over k_1^2+\kappa^2}{F(k_2^2)\over k_2^2+\kappa^2}
\biggl\{
{1\over (k_{14}+i\epsilon)} {1\over (k_{14}+k_{24}+i\epsilon)}
{1\over (k_{14}+i\epsilon)}                                     \nonumber\\
\biggr.
& &\nonumber\\
& &
\biggl.
-{1\over (k_{14}+i\epsilon)}  {1\over (k_{24}+i\epsilon)}
{1\over (k_{14}+i\epsilon)}
\biggr\}=-I_4^{(a)}(\kappa^2)
\end{eqnarray}
Here, we have taken into account the mass renormalization in the second order
(the dark disk in Fig. 4c). From the above result,
it is easy to see that in QED the contributions coming from
Fig. 4a  and Fig. 4c cancel. This is a crucial point
in solving the problem of classical limit $\hbar\to 0$ in QED \cite{EIM}
since the contribution to the self-energy of a massive fermion from the
photon vacuum polarization is proportional to

$$\Pi^{\mu\nu}_{m}(k)\propto (g^{\mu\nu} k^2-k^\mu k^\nu)
\ln\biggl[1+{k^2\over m^2}\biggr]$$
and vanishes in the infinite mass limit: $m\to\infty$.

For QCD one finds

\begin{equation}
\delta m_4^{(c)}=
2\pi\alpha_s C_F \Lambda \int\!\!{d^3\vec k\over (2\pi)^3}
{F(\vec k^2)\over \vec k^2}
\biggl[-{\alpha_s\over 4\pi} 2C_F \ln(1/\kappa^2) \biggr].
\end{equation}

\medskip
\noindent
{\bf (d) Diagram in Fig. 4d.}

The contribution from the diagrams in Fig. 4d, involving the vacuum
polarization by massless quarks, gluons, and ghost  is written as

\begin{equation}
\delta m_4^{(d)}=\Sigma_4^{(d)}(m)=
2\pi\alpha_s C_F \Lambda \int\!\! {d^3\vec k\over (2\pi)^3}
{F(\vec k^2)\over \vec k^2}\Pi^{\rm ren}(-\vec k^2)
\end{equation}

In the above expression, the renormalized vacuum polarization reads
(see Appendix for a somewhat detailed calculation)

\begin{eqnarray}
\Pi^{\rm ren}(-\vec k^2)&=&
-{\alpha_s\over 4\pi}\int\limits_0^1\!\! d\alpha
\ln\biggl[1+\alpha(1-\alpha){\vec k^2\over \kappa^2} \biggr]
\biggl\{
2N_c\biggl[1+4\alpha(1-\alpha)\biggr]-4n_f\alpha(1-\alpha)
\biggr\}\nonumber\\
& &\nonumber\\
&=&-{\alpha_s\over 4\pi}\tilde b \ln{\vec k^2\over\kappa^2}+O(1)
\end{eqnarray}
upon expanding in small parameter $\kappa^2$. Here the appearance of the
coefficient \\
$\tilde b=(10N_c-2n_f)/3$ is due to the use of the Feynman gauge.

Summing up all the contributions from the one- and two-loops, one finds

\begin{eqnarray}
\delta m&=&2\pi\alpha_s C_F \Lambda \int\!\!{d^3\vec k\over (2\pi)^3}
{F(\vec k^2)\over \vec k^2}
\biggl\{1-{\alpha_s\over 4\pi}
\biggl[N_c\ln(1/\kappa^2)+{\tilde b\over N_c}\ln(\vec k^2/\kappa^2)\biggr]
\biggr\}.
\nonumber\\
&&\\
&\to&2\pi\alpha_s C_F \Lambda \int\!\!{d^3\vec k\over (2\pi)^3}
{F(\vec k^2)\over \vec k^2}
\biggl\{1-{\alpha_s\over 4\pi}\ln(\Lambda^2/\lambda^2)
N_c\biggl[{2\over 3}+{ b\over N_c}\biggr]
\biggr\}.
\nonumber
\end{eqnarray}
Here, we express the final result in term of the first coefficient in the
Gell-Mann-Low function $b=(11N_c-2n_f)/3$. By adopting the number (flavor) of
light quarks to be three: $n_f=3$, we can see for $N_c=3$ that the
$O(\alpha_s)$ corrections  to the leading static Coulomb self-energy coming
from the bubble insertions to the gluon propagator is of the same order
of magnitude as those from other diagrams: Figs.4a-c, so the latter
contribution
should be included.

In ref. \cite{BSUV} the self-energy of the heavy quark in the heavy quark limit
was calculated by introducing the running constant $\alpha_s(\vec k^2)$ in the
Coulomb energy expression:
Eq.(9), and by introducing the explicit integration cutoff in place of the
ultraviolet form factor $F(-k^2/\Lambda^2)$ (Fig.5),

\begin{equation}
\delta m={8\pi\over 3}\int\limits_{|\vec k|<\mu_0}\!\!{d^3\vec k\over (2\pi)^3}
{\alpha_s(\vec k^2)\over \vec k^2}
\end{equation}
with

\begin{equation}
\alpha_s(\vec k^2)={\alpha_s(\mu_0^2)\over 1-(\alpha_s(\mu_0^2)b/4\pi)
\ln(\mu_0^2/\vec k^2)}.
\end{equation}
According to Ref.[3], the zero of the denominator in the above expression
defines the position of the infrared renormalon, leading to the uncertainty in
the pole mass. Here, we are interested in the $O(\alpha_s^2(\mu_0^2))$
contribution from Eq.(27),

\begin{equation}
\delta m_4=\biggl({\alpha_s\over\pi}\biggr)^2{b\over 3}
\mu_0\int\limits_0^1\!\!dt\ln{1\over t^2}
=\biggl({\alpha_s\over\pi}\biggr)^2{2b\over 3}\mu_0.
\end{equation}
Here one might regard $\mu_0$ as corresponding to our $\Lambda$ (in this
respect one could argue
that the value of the integration cutoff in Eq.(27) should be different from
the
scale $\mu_0$ introduced in Eq.(28). This would introduce an additional
contribution of a logarithmic type to Eq.(29), which would however not prohibit
the appearance of the alleged renormalon).
This result may be compared with
the part of the expression in
Eq.(26) which is proportional to $b/N_c$. Since we have not calculated the
higher loop contributions, we cannot draw any definite conclusion. However,
in view of the difference in our result: Eq.(26) and that of ref.[3],
it might be possible that due to the contributions other than the
bubble insertions in the gluon propagator the characteristics (position, etc.)
 of the alleged
infrared renormalon in the heavy quark self-energy would be different from the
original one, or even the renormalon might
not be there.

\section{Discussion}
\baselineskip 20pt
\medskip
To summarize our present note,
first we have demonstrated that the popular dimensional
regularization method yeilds only the quantum limit, thus unsuited for the
calculation of the self-energy of the heavy quark in the heavy quark
(or the classical) limit  which is just the opposite of the quantum limit.
Thus we employed next the conventional method for ultraviolet and infrared
regularizations to calculate the one- and two-loop contributions to the
self-energy of the heavy quark in the heavy quark limit. The one-loop result
is found as the standard Coulomb-like self-energy of a static colour souce
whose radius is characterized by the ultraviolet regularization mass
$\Lambda$ of the
theory. Then all the two-loop contributions have turned out to be proportional
to $\Lambda\ln(\Lambda /\lambda)$ with the infrared cutoff $\lambda$, while
the corresponding contributions in QED
internally cancel out to vanish \cite{EIM}. Of those two-loop contributions,
only the
bubble insertion to the gluon propagator in the lowrst order (Coulomb)
contribution (Fig.4d) has been discussed often. We have found that other
contributions are of the same size and have to be taken into account.

{}From our study we have reached a couple of observations (possiblly,
speculations).  The first is that since the contributions depicted in Figs.4a-c
are not very different in magnitude from the bubble insertions to the gluon
propagator at the two-loop level, the infrared renormalon in the heavy quark
self energy resulting from the
simple bubble chain summation, due to the adoption of the running coupling
constant, might be modified regarding its position, etc. or might even be
absent
once a due consideration of the remaining higher-loop effects are incorporated.

The second observation is concerned with the dimensional regularization. Our
assertion regarding the inappropriateness of in this method in the context of
the heavy quark limit should be valid also in what is called the heavy baryon
chiral perturbation theory (HBCHPT), (see for example a review \cite{MEIS}
and references therein) which combines the ordinary Chiral Perturbation theory
(CHPT)
for the octet
peudoscalar mesons coupled with baryonic (nucleons, deltas, etc.) degrees of
freedom. Here very
often the
Lagrangian (or the S-matrix) for the system is expanded in the inverse power of
the baryon mass(es), exactly like in the heavy quark expansion realized in the
heavy quark limit. The the meson loop integrals arising
from the meson-baryon
interactions are conventionally carried out using the {\it dimensional
regularization}.
Now according to our study in this note, this method is only compatible with
the opposite quantum limit ! Thus it is our feeling that the existing results
in HBCHPT should be examined again carefully from the present context.

\bigskip
\ \ \ \ \ {\large\bf Acknowlegements}
\medskip

The authors would like to acknowledge a helpful discussion with Arkady
Vainshtein. This work was supported in part by the grant DE-FG05-84ER40143
provided by the U.S. Department of Energy.

\newpage
\ \ \ \ \ {\large\bf Appendix}

\bigskip
\noindent
{\bf (a) The calculation of the integral $I^{(a)}_4(\kappa^2)$.}

Using the obvious equality in Eq. (17)

$$
{1\over k_{14}+k_{24}+i\epsilon} {1\over k_{24}+i\epsilon}=
\biggl[{1\over k_{24}+i\epsilon}-{1\over k_{14}+k_{24}+i\epsilon}\biggr]
{1\over k_{14}}
$$
and the  well-known formula

$$
{1\over k_4+i\epsilon}={P\over k_4}-i\pi\delta(k_4)
$$
one finds

$$
I_4^{(a)}(\kappa^2)={1\over 4\pi}\int\!\!{d^3\vec k\over (2\pi)^3}
{F(\vec k^2)\over \vec k^2} J(\kappa^2)
$$
where

$$
J(\kappa^2)=2\int\limits_0^\infty\!\! dt {f(0)-f(t^2)\over t^2}=
-4\int\limits_0^\infty\!\! dt f^{\prime}(t^2).
$$
We define the function $f(t^2)$ as

$$
f(t^2)=\int\!\!{d^3 \vec k\over (2\pi)^3}
{F(t^2+\vec k^2)\over t^2+\vec k^2+\kappa^2}.
$$
Using the Mellin representation: Eq.(3) for the form factor $F(k^2)$ and
integrating over $\vec k$, one can get

\begin{eqnarray}
J(\kappa^2)&=&{\sqrt{\pi}\over 2\pi^2}
{1\over 2i}\int\limits_{-\delta+i\infty}^{-\delta-i\infty}
{d\zeta\over\sin{\pi\zeta}} c(\zeta) {\Gamma(1/2-\zeta)\over \Gamma(-\zeta)}
\int\limits_0^1\!\! d\alpha (1-\alpha)^{-\zeta-1}
\int\limits_0^\infty dt [t^2+\alpha\kappa^2]^{\zeta-1/2}
\nonumber\\
&&\nonumber\\
&=&-{1\over 4\pi}
{1\over 2i}\int\limits_{-\delta+i\infty}^{-\delta-i\infty}
{d\zeta\over\sin{\pi\zeta}} c(\zeta) (\kappa^2)^{\zeta}
{\Gamma(1+\zeta)\Gamma(1-\zeta)\over \zeta}
\to {1\over 4\pi}\ln{1\over\kappa^2},
\nonumber
\end{eqnarray}

\noindent for small $\kappa$. Finally, we have

$$
I_4^{(a)}(\kappa^2)={1\over (4\pi)^2}\int\!\!{d^3\vec k\over (2\pi)^3}
{F(\vec k^2)\over \vec k^2} \ln{1\over\kappa^2} +O(1).
$$

\newpage
\noindent
{\bf (a) The calculation of $\Pi^{\rm ren}(q^2)$.}

\medskip
As an example of the calculation of diagrams in  Fig.4, we here
present the details of the calculation of the vacuum polarization
$\Pi^{\rm ren}(q^2)$ by massless gluons, ghosts, and quarks (see, insertions
to the gluon propagator in Fig.4d). We proceed in adopting
 the Feynman gauge. We shall use the dimensional regularization in
the {\it intermediate} calculations, but the final result will not depend on
the choice of the regularization.

We define the S-matrix element describing the vacuum polarization as

$$
S={i\over 2}\int\!\!\int dx_1dx_2
A^{c_1}_{\nu_1}(x_1) \Pi^{c_1c_2}_{\nu_1\nu_2}(x_1-x_2) A^{c_2}_{\nu_2}(x_2)
$$
with its Fourier transform defined as

$$
\Pi^{c_1c_2}_{\nu_1\nu_2}(q)=\int dxe^{-iqx}\Pi^{c_1c_2}_{\nu_1\nu_2}(x).
$$
For simplicity we do not introduce new notation for this value.

The contribution coming from the gluons and ghosts may be written in the form

$$
\biggl[\Pi^{c_1c_2}_{\nu_1\nu_2}(q)\biggr]_{\rm gluon+ghost}=
g^2_s f^{abc_1}f^{abc_2}
\int\limits_0^1 \! d\alpha \; \mu^{2\epsilon}\!\int {d^D k\over (2\pi)^D i}
{N^{\nu_1\nu_2}_{\rm gluon}+N^{\nu_1\nu_2}_{\rm ghost} \over
[-k^2-\alpha(1-\alpha) q^2]^2}
$$
where

\begin{eqnarray}
N^{\nu_1\nu_2}_{\rm gluon}&=&
g^{\nu_1\nu_2}(D-1)\biggl\{ {6\over D}k^2+[1-4\alpha(1-\alpha)]q^2 \biggr\}
\nonumber\\
&&\nonumber\\
&&+[q^2g^{\nu_1\nu_2}-q^{\nu_1}q^{\nu_2}]
\biggl[6-D+(4D-6)\alpha(1-\alpha)\biggr],
\nonumber\\
&&\nonumber\\
&&\nonumber\\
N^{\nu_1\nu_2}_{\rm ghost}&=&
g^{\nu_1\nu_2}\biggl\{-{2\over D}k^2+2\alpha(1-\alpha)q^2\biggr\}
-2\alpha(1-\alpha)[q^2g^{\nu_1\nu_2}-q^{\nu_1}q^{\nu_2}].
\nonumber
\end{eqnarray}

First, we show that the apparent non-gauge invariant term proportional to
$g^{\nu_1\nu_2}$ does vanish.
For this objective we use the identity:

$$
\int d^D k{ak^2+bq^2\over [-k^2-\alpha\beta q^2]^2}=
-{1\over 2\alpha\beta}\int d^D k
{D\alpha\beta a+(2-D)b\over [-k^2-\alpha\beta q^2]}
$$
which may be proved by using the following expression,

$$
{k^2\over [-k^2-\alpha\beta q^2]^2}=
k^2{d\over dk^2} {1\over [-k^2-\alpha\beta q^2]}=
{1\over 2} k^\mu {d\over dk^\mu} {1\over [-k^2-\alpha\beta q^2]},
$$
and integrating it by parts. Then the coefficient of the apparently non-gauge
invariant contribution becomes

\begin{eqnarray}
&&\int\limits_0^1 {d\alpha \over \alpha(1-\alpha)}
\int {d^D k\over [-k^2-\alpha(1-\alpha)q^2]}
\biggl[-1+4{(D-1)\over (D-2)}\alpha(1-\alpha)\biggr]
\nonumber\\
&&\nonumber\\
&=&\int d^Dk {1\over [-k^2-q^2]}
\int\limits_0^1 d\alpha [\alpha(1-\alpha)]^{D/2-2}
\biggl[-1+4{(D-1)\over (D-2)}\alpha(1-\alpha)\biggr]\nonumber\\
&&\nonumber\\
&=&\int d^Dk {1\over [-k^2-q^2]}
\biggl\{-{\Gamma^2(D/2-1)\over \Gamma(D-2)}+
4{(D-1)\over (D-2)} {\Gamma^2(D/2)\over \Gamma(D)}\biggr\}\equiv 0.
\nonumber
\end{eqnarray}
Finally one finds the gauge invariant form for the vacuum polarization
including both gluons and ghosts

$$
\biggl[\Pi^{c_1c_2}_{\nu_1\nu_2}(q)\biggr]_{\rm gluon+ghost}=
f^{abc_1}f^{abc_2}[q^2 g^{\nu_1\nu_2}-q^{\nu_1} q^{\nu_2}]
\Pi_{\rm gluon+ghost}(q^2)
$$
with $\Pi_{\rm gluon+ghost}$ being equal to

$$
\Pi_{\rm gluon+ghost}(q^2)=
\int\limits_0^1 d\alpha \mu^{2\epsilon}\int {d^D k\over (2\pi)^D i}
{[D-6-4(D-2)\alpha(1-\alpha)] \over [-k^2-\alpha(1-\alpha) q^2]^2}
$$

Note that this expression has both infrared and ultraviolet
logarithmic divergences in four dimension. To regularize
the infrared divergence we introduce the gluon (ghost) {\it mass} $\lambda$
to the denominator. Then the renormalized vacuum polarization is written as

\begin{eqnarray}
\Pi_{\rm gluon+ghost}^{\rm ren}(q^2)&=&
\Pi_{\rm gluon+ghost}(q^2)-\Pi_{\rm gluon+ghost}(0)
\nonumber\\
&&\nonumber\\
&=&-{\alpha_s\over 4\pi}\int\limits_0^1d\alpha
\ln[1-\alpha(1-\alpha){q^2\over\lambda^2}]
\biggl\{2N_c[1+4\alpha(1-\alpha)]\biggr\}
\nonumber\\
&&\nonumber\\
&=&-{\alpha_s\over 4\pi} q^2 \int\limits_{4\lambda^2}^\infty
{d m^2\over m^2}
{\sqrt{1-4\lambda^2/m^2}\over m^2-q^2}
{N_c\over 3}\biggl[10+8{\lambda^2\over m^2}\biggr].
\nonumber
\end{eqnarray}

The calculation of the vacuum polarization by quarks turns out to go along the
similar line with the result identical to the above one up to a constant
factor. The combined result is the one in Eq.(25). Note that in this equation
the momentum is measured in units of $\Lambda$.

\newpage
\listoffigures

\noindent
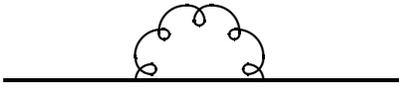
Fig. 1. The lowest order (one loop) quark self-energy diagram.

\noindent
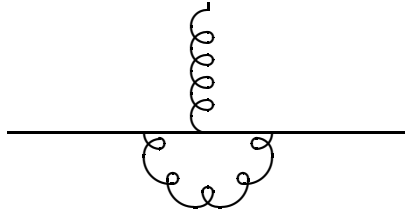
Fig. 2. An $O(\alpha_s)$ correction to the quark-gluon vertex by quark-gluon
interaction.

\noindent
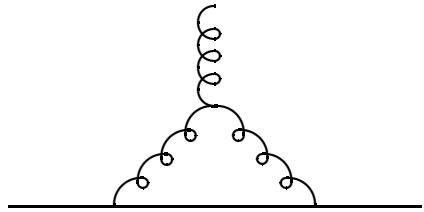
Fig. 3. an $O(\alpha_s)$ correction to the quark-gluon vertex by gluon-gluon
interaction.

\noindent
Fig. 4. Quark self-energy diagrams, to two loops. The mass counter term from
the lowest order contribution is indicated as a blob in Diagram c. In Diagram d
inserted in the gluon propagator are the gluon, ghost and light quark loops,
respectively.

\noindent
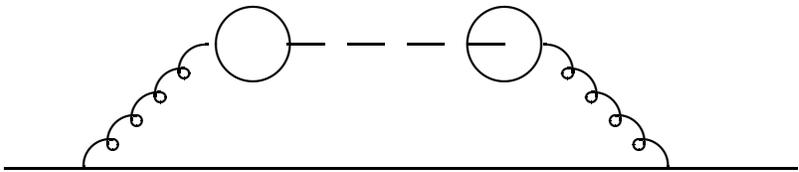
Fig. 5. Bubble chain in the gluon propagator causing the infrared renormalon,
according to Ref. 3.

\newpage
\pagestyle{empty}
\input{FEYNMAN}
\vskip 3in  \hskip -0.5in
\begin{picture}(30000,10000)
\THICKLINES
\drawline\fermion[\E\REG](2000,0)[15000]
\drawloop\gluon[\N 5](7000,0)
\put(8000,-5500){\large\bf Fig. 1}
\drawline\fermion[\E\REG](20000,0)[15000]
\drawloop\gluon[\S 5](30000,0)
\global\Xone=\pbackx   
\global\Yone=\pbacky   
\global\advance\Xone by \loopfrontx  \global\advance\Yone by \loopfronty
\global\divide\Xone by 2   \global\divide\Yone by 2
\backstemmed\drawline\gluon[\N\CENTRAL](\Xone,\Yone)[4]
\put(26000,-5500){\large\bf Fig. 2}
%
\drawvertex\gluon[\S 3](20000,-10000)[3]
\drawline\fermion[\E\REG](\vertexthreex,\vertexthreey)[4000]
\drawline\fermion[\W\REG](\vertexthreex,\vertexthreey)[4000]
\drawline\fermion[\W\REG](\vertextwox,\vertextwoy)[4000]
\drawline\fermion[\E\REG](\vertextwox,\vertextwoy)[4000]
\put(18000,-22000){\large\bf Fig. 3}
\drawline\fermion[\E\REG](2000,-30000)[15000]
\drawloop\gluon[\S 5](10000,-30000)
\global\Xone=\pbackx   
\global\Yone=\pbacky   
\global\advance\Xone by \loopfrontx  \global\advance\Yone by \loopfronty
\global\divide\Xone by 2   \global\divide\Yone by 2
\backstemmed\drawloop\gluon[\N 5](\Xone,\Yone)
\put(8000,-36000){\large\bf Fig. 4a}
\drawline\fermion[\E\REG](20000,-30000)[15000]
\drawline\gluon[\NE\REG](24000,-30000)[3]
\drawline\gluon[\SE\REG](28000,-26200)[3]
\drawline\gluon[\S\REG](28000,-26200)[3]
\put(26000,-36000){\large\bf Fig. 4b}
\end{picture}

\newpage

\vskip 3in  \hskip -0.5in
\begin{picture}(30000,10000)
\THICKLINES
\drawline\fermion[\E\REG](2000,0)[15000]
\drawline\gluon[\NE\REG](4000,0)[5]
\drawline\gluon[\SE\REG](9500,5600)[5]
\drawloop\gluon[\N 5](7200,0)
\drawline\fermion[\E\REG](20000,0)[23000]
\drawline\gluon[\NE\REG](24000,0)[4]
\drawloop\gluon[\N 10](\gluonbackx,\gluonbacky)
\drawline\gluon[\SE\REG](33500,4500)[4]
\drawline\fermion[\E\REG](2000,-10000)[15000]
\drawline\gluon[\NE\REG](4000,-10000)[5]
\drawline\gluon[\SE\REG](\gluonbackx,\gluonbacky)[5]
\put(9500,-10000){\circle*{3000}}
\put(7300,-15000){\large\bf Fig. 4c}
\drawline\fermion[\E\REG](20000,-10000)[23000]
\drawline\gluon[\NE\REG](24000,-10000)[4]
\drawline\gluon[\SE\REG](33000,-5200)[4]
\put(31000,-5000){\circle{15000}}
\drawline\fermion[\E\REG](20000,-20000)[23000]
\drawline\gluon[\NE\REG](24000,-20000)[4]
\drawline\gluon[\SE\REG](33000,-15200)[4]
\put(31000,-15000){\circle{15000}}
\put(29000,-25000){\large\bf Fig. 4d}
\drawline\fermion[\E\REG](5000,-35000)[30000]
\drawline\gluon[\NE\REG](8000,-35000)[4]
\put(14400,\gluonbacky){\circle{3000}}
\global\advance \gluonbackx by 3000
\drawline\scalar[\E\REG](\gluonbackx,\gluonbacky)[4]
\put(\scalarbackx,\scalarbacky){\circle{3000}}
\global\advance \scalarbackx by 1500
\drawline\gluon[\SE\REG](\scalarbackx,\scalarbacky)[4]
\put(17000,-39000){\large\bf Fig. 5}
\end{picture}

\begin{thebibliography}{99}
\bibitem{Tar} R.Tarrach, Nucl. Phys. {\bf B183}, 384 (1981).
\bibitem{GBGS} N.Gray, D.J.Broadhurst, W.Grafe. K.Schilcher,
Z.Phys.C {\bf C48}, 673 (1990).
\bibitem{BSUV} I.I.Bigi, M.A.Shifman, N.G.Uraltsev, A.I.Vainshtein,
Phys. Rev. {\bf D50}, 2234 (1994).
\bibitem{BB} M.Beneke and V.M.Braun, Nucl. Phys. {\bf B426}, 301 (1994);
Hep-Ph/9411229.
\bibitem{EIM} G.V.Efimov, M.A.Ivanov, O.A.Mogilevsky
Ann. of Phys. {\bf 103}, 169 (1977).
\bibitem{MEIS} U.-G. Meissner, Lecture given at Indian Summer School on
Electron
Scattering off Nucleons and Nuclei, Prague, Czech, September,
(1994);Hep-Ph/9411300.
\end{thebibliography}
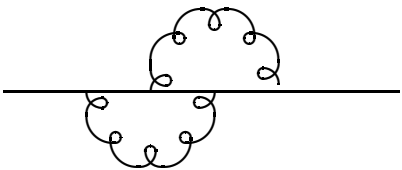
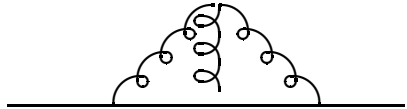
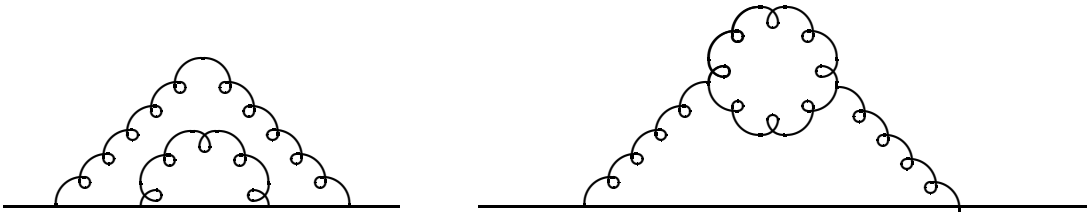
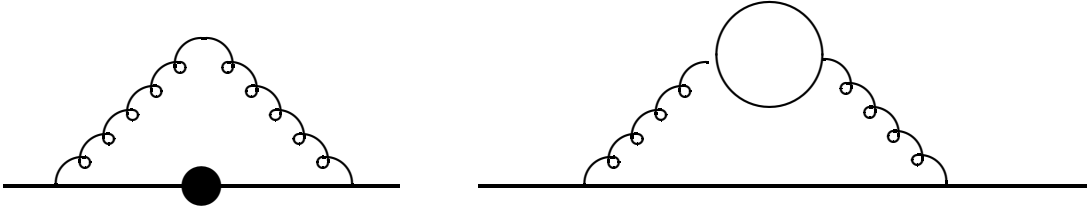
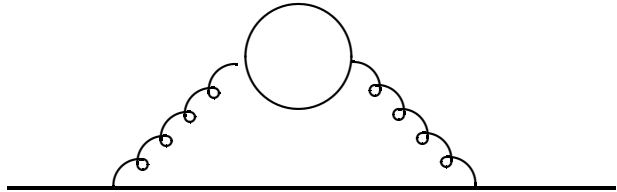
\end{document}